\documentclass[12pt]{article}%
\usepackage{graphicx}
\usepackage[english]{babel}
\usepackage{amsmath,amssymb,amsthm}

\begin{document}

\begin{center}{\large\bf Symmetries in Foundation of Quantum Theory and Mathematics} \end{center}

\vskip 1em \begin{center} {\large Felix M. Lev} \end{center}
\vskip 1em \begin{center} {\it Artwork Conversion Software Inc.,
509 N. Sepulveda Blvd, Manhattan Beach, CA 90266, USA
(Email:  felixlev314@gmail.com)} \end{center}

\begin{abstract}
In standard quantum theory, symmetry is defined in the spirit of Klein's Erlangen Program: the background space
has a symmetry group, and the basic operators should commute according to the Lie algebra of that group. We argue that the definition should be the opposite: background space has a direct physical meaning only on classical level while on quantum level symmetry should be defined by a Lie algebra of basic operators. Then the fact that de Sitter symmetry is more general than Poincare one can be proved mathematically. The problem of explaining cosmological acceleration is very difficult but,
as follows from our results, there exists a scenario that the phenomenon of cosmological acceleration can be explained proceeding from basic principles of quantum theory. The explanation has nothing to do with existence or nonexistence of dark energy and therefore the cosmological constant problem and the dark energy problem do not arise. We consider finite quantum theory (FQT) where states are elements
of a space over a finite ring or field with characteristic $p$ and operators of physical quantities act in this space. We prove that, with the
same approach to symmetry, FQT and finite mathematics are more general than standard quantum theory and classical mathematics, respectively: the latter theories are special degenerated cases of the former ones in the formal limit $p\to\infty$.
\end{abstract}

\begin{flushleft} Keywords: quantum theory, de Sitter symmetry, dark energy, finite quantum theory, finite mathematics\end{flushleft}

\section{Introduction}
\label{intro}

In standard quantum theory, symmetry is defined in the spirit of the Erlangen Program: the background space
has a symmetry group, and the basic operators for the system under consideration should commute according to the Lie algebra of
that group. According to the editorial policy of Symmetry, ``Felix Klein's Erlangen Program, and continuous symmetry``
are in the scope of the journal. The formulation of the Erlangen program was given in 1872 when quantum
theory did not exist. However, now it is clear that background space is only a classical notion and, as argued in Section \ref{fundamentaltheories}, in quantum theory the definition of symmetry should be the opposite: each system is described by a set of linearly  independent operators and, by definition, the rules how they commute with each other define the symmetry algebra. As shown in the present paper, such a change of standard paradigm 
sheds a new light on foundational problems of quantum theory and mathematics.   

Standard quantum theory is based on classical mathematics of complex Hilbert spaces and operators in them.
As shown in Section \ref{fundamentaltheories}, in the new approach to symmetry it is possible to give a 
mathematical proof that quantum theory based on de Sitter algebras is more general than quantum theory based on the Poincare algebra. As a consequence, as shown in Section 
\ref{acceleration}, there exists a scenario that the phenomenon of cosmological acceleration can be naturally explained  proceeding from basic principles of quantum theory. The explanation has nothing to do with existence or nonexistence of dark energy and therefore  the cosmological constant problem and the dark energy problem do not arise.

In Section \ref{whatmath} we argue that ultimate quantum theory should be based on finite mathematics
rather than on classical one and, as shown in the subsequent sections, the new approach to symmetry 
changes standard paradigms on what quantum theory and what mathematics are the most
general.  First, in Section \ref{infinity} we give a new look at potential vs. actual infinity.
In Section \ref{remarks} we consider standard arithmetic of natural numbers and argue that operations modulo a number
are more natural than standard arithmetic operations. In Section \ref{FQT} we prove that, as a consequence of the proof given in Section \ref{proof}, finite quantum theory and finite mathematics are more general
than standard quantum theory and classical mathematics, respectively: the latter theories are special
degenerated cases of the former ones in the formal limit when the characteristic of the ring or field in the
former theories goes to infinity. Finally Section \ref{conclusion} includes the discussion.

\section{Comparison of different physical and mathematical theories}
\label{fundamentaltheories}

When two physical or mathematical theories are compared
one can pose a problem whether one of them is more general than the other.

One of the known examples is the comparison 
of nonrelativistic theory (NT) with relativistic one (RT). 
 Usually RT is treated as more general than NT for several reasons. The basic principles of RT are treated
as more general than those for NT but a necessary condition for theory A to be treated as more
general than theory B is that A should describe experiment better than B, and indeed RT
describes experiment better than NT. A problem arises whether there are  mathematical arguments in favor of the statement that RT is more general than NT. One of the arguments is that RT contains a finite parameter $c$ and NT can be treated as a special degenerated
case of RT in the formal limit $c\to\infty$. Therefore, by choosing a large value of $c$,  RT can reproduce any result of NT with a high accuracy. On the contrary, when the limit is already taken one cannot return back from NT to RT and NT cannot reproduce all results of RT. It can reproduce only results obtained when $v\ll c$. 

Other known examples
are that classical theory is a special degenerated case of quantum one in the formal limit $\hbar\to 0$ and
RT is a special degenerated case of de Sitter (dS) and anti-de Sitter (AdS) invariant theories in the formal
limit $R\to\infty$ where $R$ is the parameter of contraction from the dS or AdS algebras to the Poincare algebra (see below). A problem arises whether it is possible to give a general definition when theory A is more
general than theory B. In view of the above examples, we propose the following

{\bf Definition:} {\it Let theory A contain a finite parameter and theory B be obtained from theory A in the formal limit when the parameter goes to zero or infinity. Suppose that with any desired accuracy theory A can reproduce any result of theory B by choosing a value of the parameter. On the contrary, when the limit is already
taken then one cannot return back to theory A, and theory B cannot reproduce all results of theory A. Then theory A is more general than theory B and theory B is a special degenerated case of theory A}. 
A problem arises how to justify this
{\bf Definition} not only from physical but also from mathematical considerations.

In relativistic quantum theory the usual approach to symmetry on quantum level follows. 
Since the Poincare group is the group of motions of Minkowski space, quantum states should be described by representations of this group. 
This implies that the representation generators acting in the Hilbert space of the system under consideration commute according to the commutation relations of the Poincare group Lie algebra:
\begin{eqnarray}
&&[P^{\mu},P^{\nu}]=0,\quad [P^{\mu},M^{\nu\rho}]=-i(\eta^{\mu\rho}P^{\nu}-
\eta^{\mu\nu}P^{\rho}),\nonumber\\
&&[M^{\mu\nu},M^{\rho\sigma}]=-i (\eta^{\mu\rho}M^{\nu\sigma}+\eta^{\nu\sigma}M^{\mu\rho}-
\eta^{\mu\sigma}M^{\nu\rho}-\eta^{\nu\rho}M^{\mu\sigma})
\label{PCR}
\end{eqnarray}
where $\mu,\nu=0,1,2,3$, $P^{\mu}$ are the operators of the four-momentum and  $M^{\mu\nu}$ are the operators of Lorentz angular momenta. This approach is in the spirit of Klein's Erlangen Program in mathematics.

However, background space has a direct physical meaning only on classical level, and this is clear even from the
fact that in quantum theory neither time nor coordinates can be measured with the absolute accuracy
(see a more detailed discussion below). In QED, QCD and electroweak theory the Lagrangian density 
depends on the four-vector $x$ which is associated with a point in Minkowski space but this is only the
integration parameter which is used in the intermediate stage. The goal of the theory is to construct the
$S$-matrix and when the theory is already constructed
one can forget about Minkowski space because no physical quantity depends on $x$. This is in the
spirit of the Heisenberg $S$-matrix program according to which in relativistic quantum theory it is possible to describe only transitions of states from the infinite past when $t\to -\infty$ to the distant future 
when $t\to +\infty$. 

Note that the fact that the $S$-matrix is the operator in momentum space does not exclude a possibility that
in some situations it is possible to have a space-time description with some accuracy but not with absolute accuracy. First of all, the problem of time is one of
the most important unsolved problems of quantum theory (see e.g. Reference \cite{time} and references therein), and time cannot be measured with the accuracy better than $10^{-18}s$. Also, in typical situations the position operator in momentum representation exists not only
in the nonrelativistic case but in the relativistic case as well. In the latter case it is known, for example, as
the Newton-Wigner position operator \cite{NW} or its modification (see e.g. Reference \cite{position}). As pointed
out even in textbooks on quantum theory, the coordinate description of elementary particles can work only in some approximations. In particular, even in most favorable scenarios, for a massive particle with the mass $m$ its coordinate cannot be measured with the accuracy better than the particle Compton wave length $\hbar/mc$ \cite{BLP}.

For illustration of the background problem consider first classical electrodynamics. We know that the electromagnetic field 
consists of photons but on classical level the theory does not describe the state of each photon. The classical
electromagnetic fields ${\bf E}({\bf r},t)$ and ${\bf B}({\bf r},t)$ describe the effective contribution
of all photons at the point $x=({\bf r},t)$ of Minkowski space, and in classical (non-quantum) theory it is
assumed that the parameters $({\bf r},t)$ can be measured with any desired accuracy.

On quantum level a problem arises how to define the photon coordinate wave function. For example,
a section in the known textbook \cite{AB} is titled "Impossibility of introducing the
photon wave function in coordinate representation". On the other hand, a detailed discussion of the photon position 
operator in papers by Margaret Hawton \cite{Hawton,Hawton1,Hawton2} and references therein indicates that it is
possible to define the photon coordinate wave function $\psi({\bf r},t)$ but the description with such a
wave function can have a good accuracy only in semiclassical approximation (see also Reference \cite{position}), and coordinates cannot be
directly measured with the accuracy better than the size of the hydrogen atom. 

In particle physics distances are never
measured directly, and the phrase that the physics of some process is defined by characteristic distances $l$ means
only that if $q$ is a characteristic momentum transfer in this process then $l=\hbar/q$. This conclusion is based
on the assumption that coordinate and momentum representations in quantum theory are related to each other by the
Fourier transform. However, as shown in Reference \cite{position}, this assumption is based neither on strong theoretical 
arguments nor on experimental data.

Local quantum field theories (QFT) work with local quantized field operators $\varphi(x)$. Such operators act in the Fock
space of the system under consideration. Here the quantity $x$ is not related to any particle, this is only a formal parameter. One of the principles of quantum theory is that any physical quantity should be described by an
operator. However, since $x$ is not related to any particle, there is no operator related to $x$. Therefore
$x$ cannot be directly measured and $\varphi(x)$ does not have a direct physical meaning. Strictly speaking,
even the word "local" here might be misleading since $x$ is not related to any particle. 

Foundational problems of QFT have been discussed by many authors. One of the main problems in substantiating QFT is that QFT contains products of interacting local quantized fields at the same points. 
As explained in textbooks (e.g. in the book \cite{Bogolubov}), such fields can be treated only as distributions, and the product of distributions at the same point is not a correct mathematical operation. As a consequence, in QFT there are divergences and other inconsistencies. It is rather strange that many physicists believe that such products are needed to preserve locality. However, such products have nothing to do with locality because $x$ is not a physical quantity. 

As stated in the introductory section of the textbook \cite{BLP}, local quantum fields and Lagrangians are rudimentary notions which will disappear in the ultimate quantum theory. My observation is that now physicists usually do not believe that such words could be written in such a known textbook. The reason is that in view of successes of QCD and electroweak theory for explaining experimental data those ideas have become almost forgotten. However, although the successes are rather impressive, they do not contribute to resolving inconsistencies in QFT. Also, in the textbook 
\cite{Bogolubov} devoted to mathematical aspects of QFT, products of interacting quantum local fields are never used.

In QED one can formally define 
the operators ${\bf E}(x)$ and ${\bf B}(x)$ which are local quantized
field operators acting in the Fock space for the quantum electromagnetic field. 
However, since $x$ is not related to any photon, those operators do not define observable
physical quantities. Those operators are used in theory such that integrals of their combinations over 
space-like hypersurfaces of Minkowski space define the energy-momentum and angular momentum operators
of the electromagnetic field. So the situation is similar to that mentioned above when $x$ in the Lagrangian
density is only the integration parameter.

For illustration of the foundational problems of QFT, consider a photon emitted in the famous
21$cm$ transition line between the hyperfine energy levels of the hydrogen atom. The phrase that the lifetime of 
this transition is of the order of $\tau=10^7$ years is understood such that the width of the level 
is of the order of $\hbar/\tau$ i.e. the uncertainty of the photon energy is $\hbar/\tau$. 
In this situation a description of the system (atom + electric
field) by the wave function (e.g. in the Fock space) depending on a continuous parameter $t$ has no physical meaning 
(since roughly speaking the quantum of time in this process is of the order of $10^7$ years). 

The above examples show that space-time description with continuous space-time parameters cannot
be universal in all situations. In particular, the notion of background space cannot be universal on quantum level.
For all those reasons, as argued in Reference \cite{PRD}, the approach to symmetry should be the opposite to that
proceeding from the Erlangen Program. Each system is described by a set of linearly  independent operators.
By definition, the rules how they commute with each other define the symmetry algebra. 
In particular, {\it by definition}, Poincare symmetry on quantum level means that the operators commute
according to Eq. (\ref{PCR}). This definition does not involve Minkowski space at all.

Such a definition of symmetry on quantum level has been proposed in Reference \cite{BKT} and in
subsequent publications of those authors. I am very grateful to Leonid Avksent'evich Kondratyuk for explaining me this definition during our collaboration. I believe that this replacement of standard paradigm is fundamental for understanding quantum theory, and I did not succeed in finding a similar idea in the literature. 

Our goal is to compare four theories: classical (i.e. non-quantum) theory, nonrelativistic quantum theory, relativistic quantum theory and de Sitter (dS) or anti-de Sitter (AdS) quantum theory. All those theories are described by representations of the symmetry 
algebra containing ten linearly independent operators $A_{\alpha}\,\, (\alpha=1,2,...10)$: four energy-momentum operators, three angular momentum operators and three Galilei or Lorentz boost operators. For definiteness we assume that the 
operators $A_{\alpha}$ where $\alpha=1,2,3,4$ refer to energy-momentum operators, the operators $A_{\alpha}$ where $\alpha=5,6,7$ refer to angular momentum operators and the operators $A_{\alpha}$ where $\alpha=8,9,10$ refer to Galilei or 
Lorentz boost operators. Let $[A_{\alpha},A_{\beta}]=ic_{\alpha\beta\gamma}A_{\gamma}$ where summation over repeated indices is assumed. In the theory of Lie algebras the quantities $c_{\alpha\beta\gamma}$ are called the structure constants. 

Let $S_0$ be a set 
of $(\alpha,\beta)$ pairs such that $c_{\alpha\beta\gamma}=0$ for all values of $\gamma$ and $S_1$ be a set of $(\alpha,\beta)$ pairs such that $c_{\alpha\beta\gamma}\neq 0$ at least for some values of $\gamma$. Since
$c_{\alpha\beta\gamma}=-c_{\beta\alpha\gamma}$ it suffices to consider only such $(\alpha,\beta)$ pairs
where $\alpha<\beta$. If $(\alpha,\beta)\in S_0$ then the operators $A_{\alpha}$ and $A_{\beta}$ commute
while if $(\alpha,\beta)\in S_1$ then they do not commute.
Let $(S_0^A,S_1^A)$ be the sets $(S_0,S_1)$ for theory A and $(S_0^B,S_1^B)$ be the sets $(S_0,S_1)$ for theory B. As noted above, we will consider only theories where $\alpha,\beta=1,2,...10$. Then one can prove the following 

{\bf Statement:} {\it Let theory A contain a finite parameter and theory B be obtained from theory A in the formal limit when the parameter goes to zero or infinity. If the sets $S_0^A$ and $S_0^B$ are different and $S_0^A 
\subset S_0^B$ (what equivalent to $S_1^B \subset S_1^A$)
then theory A is more general than theory B and theory B is a special degenerated case of theory A.}

Proof: Let ${\tilde S}$ be the set of $(\alpha,\beta)$ pairs such that $(\alpha,\beta)\in S_1^A$ and
$(\alpha,\beta)\in S_0^B$. Then, in theory B, $c_{\alpha\beta\gamma}=0$ for any $\gamma$. One can choose
the parameter such that in theory A all the quantities $c_{\alpha\beta\gamma}$ are arbitrarily small.
Therefore, by choosing a value of the parameter, theory A can reproduce any result of theory B with any
desired accuracy. When the limit is already taken then, in theory B, $[A_{\alpha},A_{\beta}]=0$ for all
$(\alpha,\beta)\in {\tilde S}$. For a set of mutually commuting selfadjourned operators there
exists a basis such that its basis elements are eigenvectors of all the operators from the set.
This means that the operators $A_{\alpha}$ and $A_{\beta}$ become
fully independent and therefore there is no way to return to the situation when they do not commute.
Therefore for theories A and B the conditions of {\bf Definition} are satisfied.

It is sometimes stated that the expressions in Eq. (\ref{PCR}) are written in the system of units $c=\hbar =1$. Strictly speaking, this statement is not correct because
for the construction of relativistic quantum theory based on those equations neither $c$ nor $\hbar$ is needed. 
The notion of the system of units is purely
classical, and a problem arises whether quantum theory should involve this notion. 
For example, in the Copenhagen interpretation of quantum theory, measurement is treated as an interaction with
a classical object. Therefore, in this interpretation, quantum theory is not fully independent on classical
one. On the other hand, the Copenhagen interpretation probably cannot be universal. For example,
according to the present knowledge, at the very early stages of the Universe classical objects did not
exist. 

In the representation (\ref{PCR}) the operators $M^{\mu\nu}$ are dimensionless. However, this theory still depends on systems of units because the operators $P^{\mu}$ have the dimension $length^{-1}$.
In particular, standard angular momentum
operators $(J_x,J_y,J_z)=(M^{12},M^{31},M^{23})$ are dimensionless and satisfy the commutation relations
\begin{equation}
[J_x,J_y]=iJ_z,\quad [J_z,J_x]=iJ_y,\quad [J_y,J_z]=iJ_x
\label{J}
\end{equation}
For comparison with classical theory, all physical quantities in both theories should be expressed in the
same units. For this reason one can impose a requirement that the operators $M^{\mu\nu}$
should have the dimension $kg\cdot m^2/s$. Then they should be replaced by $M^{\mu\nu}/\hbar$, respectively.
In that case the new commutation relations will have the same form as in Eqs. (\ref{PCR}) and (\ref{J}) but
the right-hand-sides will contain the additional factor $\hbar$.

As shown in quantum theory, in the representation (\ref{J}) the results for angular momenta are given by 
half-integer numbers
$0, \pm 1/2, \pm 1,...$. One can say that in units where the angular momentum is a half-integer $l$, its
value in $kg\cdot m^2/s$ is $1.0545718\cdot 10^{-34}\cdot
l\cdot kg\cdot  m^2/s$. Which of those two values has more
physical significance? In units where the angular momentum
components are half-integers, the commutation relations (\ref{J})
do not depend on any parameters. Then the meaning of
$l$ is clear: it shows how large the angular momentum is in
comparison with the minimum nonzero value 1/2. At the same time,
the measurement of the angular momentum in units $kg\cdot
m^2/s$ reflects only a historic fact that at macroscopic
conditions on the Earth between the 18th and 21st
centuries people measured the angular momentum in such units.

We conclude that for quantum theory itself the quantity $\hbar$ is not needed. However, it
is needed for the transition from quantum theory to classical one: we introduce $\hbar$, then 
the operators $M^{\mu\nu}$ have the dimension $kg\cdot m^2/s$, and since the right-hand-sides
of Eqs. (\ref{PCR}) and (\ref{J}) in this case contain an additional factor $\hbar$, all the 
commutation relations disappear in the formal limit $\hbar\to 0$. Therefore in classical theory
the set $S_1$ is empty and all the $(\alpha,\beta)$ pairs belong to $S_0$. Since in quantum theory
there exist $(\alpha,\beta)$ pairs such that the operators $A_{\alpha}$ and $A_{\beta}$ do not commute
then in quantum theory the set $S_1$ is not empty and, as follows from {\bf Statement},
classical theory is a special degenerated case of quantum one in the formal limit $\hbar\to 0$.
Since in classical theory all operators commute with each other then in this theory operators are not needed
and one can work only with physical quantities. A question why $\hbar$ is as is does not arise
since the answer is: because people's choice
is to measure angular momenta in $kg\cdot m^2/s$.

Consider now the relation between RT and NT. If we introduce the Lorentz boost operators 
$L^j=M^{0j}\,\, (j=1,2,3)$ then Eqs. (\ref{PCR}) can be written as
\begin{eqnarray}
&&[P^0,P^j]=0,\quad [P^j,P^k]=0, \quad [J^j,P^0]=0,\quad [J^j,P^k]=
i\epsilon_{jkl}P^l,\nonumber\\
&&[J^j,J^k]=i\epsilon_{jkl}J^l,\quad [J^j,L^k]=i\epsilon_{jkl}L^l,\quad [L^j,P^0]=iP^j
\label{RT1}
\end{eqnarray}
\begin{equation}
[L^j,P^k]=i\delta_{jk}P^0,\quad [L^j,L^k]=-i\epsilon_{jkl}J^l
\label{RT2}
\end{equation}
where $j,k,l=1,2,3$, $\epsilon_{jkl}$ is the fully asymmetric tensor such that $\epsilon_{123}=1$, 
$\delta_{jk}$ is the Kronecker symbol and a summation over repeated indices is assumed.
If we now define the energy and Galilei boost operators as $E=P^0c$ and $G^j=L^j/c\,\, (j=1,2,3)$,
respectively then the new expressions in Eqs. (\ref{RT1}) will have the same form while instead of
Eq. (\ref{RT2}) we will have
\begin{equation}
[G^j,P^k]=i\delta_{jk}E/c^2,\quad [G^j,G^k]=-i\epsilon_{jkl}J^l/c^2
\label{NT2}
\end{equation}

Note that for relativistic theory itself the quantity $c$ is not needed. In this theory the primary quantities
describing particles are their momenta ${\bf p}$ and energies $E$ while the velocity ${\bf v}$ of a
particle is {\it defined} as ${\bf v}={\bf p}/E$. This definition does not involve meters and seconds, and
the velocities ${\bf v}$ are dimensionless quantities such that 
$|{\bf v}|\leq 1$ if tachyons are not
taken into account. One needs $c$ only for having a possibility to compare RT and NT: when we introduce $c$ then the 
velocity of a particle becomes ${\bf p}c^2/E$, and its dimension becomes $m/s$. In this case, 
instead of the operators $P^0$ and $L^j$ we work with the operators $E$
and $G^j$, respectively. If $M$ is the Casimir operator for the Poincare algebra defined such that
$M^2c^4=E^2-{\bf P}^2c^2$ then in the formal limit $c\to\infty$ the first expression in Eq. (\ref{NT2})
becomes $[G^j,P^k]=i\delta_{jk}M$ while the commutators in the second expression become zero.
Therefore in NT the $(\alpha,\beta)$ pairs with $\alpha,\beta=8,9,10$ belong to $S_0$ while in RT
they belong to $S_1$. Therefore, as follows from {\bf Statement}, NT is a special degenerated case of RT in 
the formal limit $c\to\infty$. The question why $c=3\cdot 10^8 m/s$ and not, say
$c=7\cdot 10^9 m/s$ does not arise since the answer is: because people's choice
is to measure velocities in $m/s$.

From the mathematical point of view, $c$ is the parameter of contraction from the Poincare algebra to the Galilei one. This
parameter must be finite: the formal case $c=\infty$ corresponds to the situation when the Poincare algebra does not exist because it becomes the Galilei algebra.

In his famous paper "Missed Opportunities" \cite{Dyson} Dyson notes that RT is more general than NT, and dS and AdS theories
are more general than RT not only from physical but also from pure mathematical considerations. Poincare group is
more symmetric than Galilei one and the transition from the former to the latter at $c\to\infty$ is called contraction. Analogously dS and AdS groups are more symmetric than Poincare one and the transition from the former to the latter at $R\to\infty$ (described below) also is called contraction. At the same time, since dS and AdS groups are semisimple they have a maximum possible symmetry and cannot be obtained from more symmetric groups by contraction.
However, since we treat symmetry not from the point of view of a group of motions for the corresponding background space but from the point of view of commutation relations in the symmetry algebra, we will discuss 
the relations between the dS and AdS algebra on one hand and the Poincare algebra on the other.

By analogy with the definition of Poincare symmetry on quantum level, the definition of dS symmetry on quantum level should not
involve the fact that the dS group is the group of motions of dS space.
Instead, {\it the definition} is that the operators $M^{ab}$ ($a,b=0,1,2,3,4$, $M^{ab}=-M^{ba}$)
describing the system under consideration satisfy the
commutation relations {\it of the dS Lie algebra} so(1,4), {\it i.e.}
\begin{equation}
[M^{ab},M^{cd}]=-i (\eta^{ac}M^{bd}+\eta^{bd}M^{ac}-
\eta^{ad}M^{bc}-\eta^{bc}M^{ad})
\label{CR}
\end{equation}
where $\eta^{ab}$ is the diagonal metric tensor such that
$\eta^{00}=-\eta^{11}=-\eta^{22}=-\eta^{33}=-\eta^{44}=1$.
The {\it definition} of AdS symmetry on quantum level is given by the same equations
but $\eta^{44}=1$.

With such a definition of symmetry on quantum level, dS and AdS
symmetries are more natural than Poincare symmetry. In the
dS and AdS cases all the ten representation operators of the symmetry
algebra are angular momenta while in the Poincare case only six
of them are angular momenta and the remaining four operators
represent standard energy and momentum. In the representation (\ref{CR}) all the operators are
dimensionless, and the theory does not depend on the system of units. If we {\it define} the
operators $P^{\mu}$ as $P^{\mu}=M^{4\mu}/R$ where $R$ is a parameter with the dimension
$length$ then in the formal
limit when $R\to\infty$, $M^{4\mu}\to\infty$ but the quantities
$P^{\mu}$ are finite, Eqs. (\ref{CR}) become Eqs. (\ref{PCR}). This procedure is called contraction and 
in the given case it is the same for the dS or AdS symmetries. As follows from Eqs. (\ref{PCR}) and
(\ref{CR}), if $\alpha,\beta=1,2,3,4$ then the $(\alpha,\beta)$ pairs belong to $S_0$ in
RT and to $S_1$ in dS and AdS theories. Therefore, as follows from {\bf Statement}, 
RT is indeed
a special degenerated case of dS and AdS theories in the formal limit $R\to\infty$. 
By analogy with the abovementioned fact that $c$ must be finite, $R$ must be finite too: the formal case $R=\infty$ corresponds to the 
situation when the dS and AdS algebras do not exist because they become the Poincare algebra.

Note that the operators in Eq. (\ref{CR}) do not depend on $R$ at all. This
quantity is needed only for transition from dS quantum theory to
Poincare quantum theory. Although $R$ has the dimension $length$, it has nothing to do with the radius of the background space which,
as noted above, has a direct physical meaning only in classical theory: $R$ is simply the coefficient of proportionality between
the operators $M^{4\mu}$ and $P^{\mu}$.  
In full analogy with the above discussion of the quantities $\hbar$ and $c$, 
a question why $R$ is as is does not arise and the answer is: because people's choice is to measure distances in meters.

We have proved that all the three discussed comparisons satisfy the conditions formulated in {\bf Definition} above.
Namely, the more general theory contains a finite
parameter which is introduced for having a possibility to compare this theory with a less general one.
Then the less general theory can be treated as a special degenerated case of the former in the
formal limit when the parameter goes to zero or infinity. The more general theory can reproduce all results of
the less general one by choosing some value of the parameter. On the contrary, when the limit is already taken
one cannot return back from the less general theory to the more general one. 

In References \cite{JPA,Symm1} we considered properties of dS quantum theory and argued that
dS symmetry is more natural than Poincare one. However, the above discussion proves that dS
and AdS symmetries are not only more natural than Poincare symmetry but more general.
In particular, $R$ is fundamental to the same extent as $\hbar$ and $c$ and, as noted above, $R$ {\bf must be finite}.
Indeed, $\hbar$ is the contraction parameter from quantum Lie algebra to the classical one, $c$ is the
contraction parameter from Poincare invariant quantum theory to Galilei invariant quantum theory, and $R$ is
the contraction parameter from dS and AdS quantum theories to Poincare invariant quantum theory. 

\section{Cosmological acceleration as a consequence of quantum dS symmetry}
\label{acceleration}

The goal of this section is to prove that the above results can be applied for the solution
of the famous problem of cosmological acceleration. In Subsec. \ref{CC} we describe the cosmological constant and dark energy problems and Subsec. \ref{2bodies} gives an explanation of  cosmological acceleration problem.

\subsection{Brief overview of the cosmological constant and dark energy problems}
\label{CC}

The history of General Relativity (GR) is described in a vast literature. The Lagrangian of GR is linear in Riemannian curvature $R_c$, but from the point of view of symmetry requirements there exist infinitely many
Lagrangians satisfying such requirements. For example, $f(R_c)$ theories of gravity are widely discussed, where there can be many possibilities for choosing the function $f$.
 Then the effective gravitational
constant $G_{eff}$ can considerably differ from standard gravitational constant $G$. It is also argued that GR is a low energy approximation of more general theories involving higher order derivatives. The nature of gravity on quantum level is a problem, and standard canonical quantum gravity is not renormalizable. For those reasons
the quantity $G$ can be treated only as a phenomenological parameter but not fundamental
one.

Let us first consider the cosmological constant and dark energy problems in the framework of standard GR. Here the Einstein equations depend on two arbitrary parameters $G$ and $\Lambda$ where $\Lambda$ is the cosmological constant (CC). In the formal limit when matter disappears, space-time becomes Minkowski space when $\Lambda=0$, dS space when $\Lambda>0$, and AdS space when $\Lambda<0$.

Known historical facts are
that first Einstein included $\Lambda$ because he believed that the Universe should be stationary, and this is possible only if $\Lambda \neq 0$. However, according to Gamow,  after Friedman's results and Hubble's discovery of the Universe
expansion, Einstein changed his mind and said that inclusion of $\Lambda$ was the greatest blunder of his life. 

The usual philosophy of GR is that curvature is created by matter and therefore
$\Lambda$ should be equal to zero. This philosophy had been advocated even
in standard textbooks written before 1998. For example, the authors of Reference \cite{LL}
say that "...there are no convincing reasons, observational and theoretical, for introducing
a nonzero value of $\Lambda$" and that "... introducing to the density of the Lagrange function a constant term which does not depend on the field state would mean attributing to space-time a principally ineradicable curvature which is related neither to matter nor to gravitational waves".

However, the data of Reference \cite{Perlmutter} on supernovae have shown that $\Lambda > 0$ with the accuracy better than $5\%$, and further investigations have improved the accuracy to $1\%$. For reconciling this fact with the philosophy of GR, the terms with 
$\Lambda$ in the left-hand-sides of the Einstein equations have been moved to the 
right-hand-sides and interpreted not as the curvature of empty space-time but as a
contribution of unknown matter called dark energy. Then, as follows from the experimental value of $\Lambda$, dark energy contains approximately $70\%$ of the
energy of the Universe. At present a possible nature of dark energy is discussed in a
vast literature and several experiments have been proposed (see e.g. the review papers \cite{PR,Cai}
and references therein).

Let us note the following. In the formalism of GR the coordinates
and curvature are needed for the description of real bodies. One of fundamental principles of physics is that definition of a physical quantity is the description on how this quantity should be measured. In the Copenhagen formulation of quantum theory measurement is an interaction with a classical object. Therefore since in empty space-time nothing can be measured, the coordinates and
curvature of empty space-time have no physical meaning. This poses a problem whether
the formal limit of GR when matter disappears but space-time remains is physical. Some authors (see e.g. Reference \cite{Hikin}) propose approaches such that if matter disappears then space-time disappears too. 

The CC problem is as follows. In standard QFT one starts from the choice of the space-time background. By analogy with the philosophy of GR,
it is believed that the choice of the Minkowski background is more physical than the choice of the dS or AdS one. Here the quantity $G$ is treated as fundamental and 
the value of $\Lambda$ should be extracted from the vacuum expectation value of
the energy-momentum tensor. The theory contains strong
divergences and a reasonable cutoff gives for $\Lambda$ a value exceeding
the experimental one by 120 orders of magnitude. This result is expected because in
units $c=\hbar=1$ the dimension of $G$ is $m^2$, the dimension of $\Lambda$ is
$m^{-2}$ and therefore one might think than $\Lambda$ is of the order of $1/ G$
what exceeds the experimental value by 120 orders of magnitude. 

Several authors argue that the CC problem does not exists. For example, the
authors of Reference \cite{Bianchi} titled "Why all These Prejudices Against a Constant?"
note that since the solution of the Einstein 
equations depends on two {\it arbitrary phenomenological}  constants $G$ and 
$\Lambda$ it is not clear why we should choose only a special case $\Lambda=0$.  
If $\Lambda$ is as small as given in Reference \cite{Perlmutter} then
it has no effect on the data in Solar System and the contribution of $\Lambda$ is
important only at cosmological distances. Also theorists supporting Loop Quantum Gravity
say that the preferable choice of Minkowski background contradicts the background
independence principle. Nevertheless, majority of physicists working in this
field believe that the CC problem does exist and the solution should be sought in
the framework of dark energy, quintessence and other approaches. 

One of the consequences of the results of Section \ref{fundamentaltheories} is that the CC problem does not exist
because its formulation is based on the incorrect assumption that RT is more general  than
dS and AdS theories. Note that the operators in Eq. (\ref{CR}) do not depend on $R$ at all. As noted in 
Section \ref{fundamentaltheories}, this
quantity is needed only for transition from dS quantum theory to
Poincare quantum theory, and $R$ has nothing to do with the radius of background space and with the
cosmological constant. Also,
as noted in Section \ref{fundamentaltheories}, in full analogy with the above discussion of the quantities $\hbar$ and $c$, 
a question why $R$ is
as is does not arise and the answer is: because people's choice
is to measure distances in meters.

On the other hand, there exists a wide literature where standard GR is modified and,
as a consequence, the existing data can be explained by using alternatives of dark energy
(see e.g. References \cite{NO1,NO2,NOS,NOO} and references therein). Also, as it has been pointed out
in Reference \cite{Maxim}, the existing cosmological data can be explained only in approaches
extending Standard Model. On the other hand, as stated in Reference \cite{Deur}, the data
can be explained in the framework of existing theories.

In this section we will not discuss what approaches better describe the existing data. 
The goal of the section is limited to the consideration of the pure mathematical problem: we consider
quantum theory where dS symmetry is understood as explained in the preceding section, i.e.
that the basic operators of the theory commute according to Eq. (\ref{CR}). Then, as shown in
Subsec. \ref{2bodies}, there necessarily exists a relative
acceleration in the system of {\it free} bodies, and in semiclassical approximation the acceleration
is given by the same formula as in GR if the radius of background space equals $R$ and
$\Lambda=3/R^2$.

Before proceeding to the derivation we would like to note the following.
On classical level,  dS space is usually treated as the
four-dimensional hypersphere in the five-dimensional space such that
\begin{equation}
x_1^2+x_2^2+x_3^2+x_4^2-x_0^2=R^{'2}
\label{dSspace}
\end{equation}
where $R'$ is the radius of dS space and at this stage it is not clear whether or 
not $R'$ coincide with $R$. 
Transformations from the dS group are usual and hyperbolic rotations of this
space. They can be parametrized by usual and hyperbolic angles and do not
depend on $R'$. In particular, if instead of $x_a$ we introduce the quantities
$\xi_a=x_a/R'$ then the dS space can be represented as a set of points 
\begin{equation}
\xi_1^2+\xi_2^2+\xi_3^2+\xi_4^2-\xi_0^2=1
\label{dSxispace}
\end{equation}
Therefore in classical dS theory itself the quantity $R'$ is not needed at all. It is needed
only for transition from dS space to Minkowski one: we choose $R'$ in meters,
then  the curvature of this space is 
$\Lambda=3/R^{'2}$ and a vicinity of the point $x_4=R'$ or
$x_4=-R'$ becomes Minkowski space in the formal limit $R'\to\infty$.
Analogous remarks are  valid for the transition from AdS theory to Poincare one,
and in this case $\Lambda=-3/R^{'2}$.

\subsection{A system of two bodies in quantum dS theory}
\label{2bodies}

Let us stress that the above proof
that dS symmetry is more general than Poincare one has been performed on pure quantum
level. In particular, the proof does not involve the notion of background space 
and
the notion of $\Lambda$. Therefore a problem arises whether this result can be used for
explaining that experimental data can be described in the framework of GR with $\Lambda>0$.

Our goal is to show that in quantum mechanics based on the dS algebra, classical equations of 
motions for a system of two free macroscopic bodies follow from quantum mechanics in
semiclassical approximation, and those equations are the same as in GR with dS background space. 
We will assume that distances between the
bodies are much greater than their sizes, and the bodies do not
have anomalously large internal angular momenta. Then, from the formal
point of view, the motion of two bodies as a whole can be described by the same formulas as the motion of two elementary particles with zero spin. 
In quantum dS theory elementary particles are described by irreducible representations (IRs) of the dS algebra and, as shown in References \cite{JPA,Symm1},  one can explicitly construct such IRs. 

It is known that in Poincare theory any massive IR can be implemented in the 
Hilbert space of functions $\chi({\bf v})$ on
the Lorenz 4-velocity hyperboloid with the points $v=(v_0,{\bf v}),\,\, v_0=(1+{\bf v}^2)^{1/2}$ such that 
$\int\nolimits |\chi({\bf v})|^2d\rho({\bf v}) <\infty$ and $d\rho({\bf v})=d^3{\bf v}/v_0$ is the Lorenz
invariant volume element. For positive energy IRs the value of energy is $E=mv_0$
where $m$ is the particle mass {\it defined as the positive square root} $(E^2-{\bf P}^2)^{1/2}$.
Therefore for massive IRs, $m>0$ by definition.

However, as shown by Mensky \cite{Mensky}, in contrast to Poincare theory, IRs in dS theory can be implemented only on two Lorenz hyperboloids,
i.e. Hilbert spaces for such IRs consist of sets of two functions $(\chi_1({\bf v}),\chi_2({\bf v}))$ such that 
$$\int\nolimits (|\chi_1({\bf v})|^2+|\chi_2({\bf v})|^2)d\rho({\bf v}) <\infty$$
In Poincare limit one dS IR splits into two IRs of the Poincare algebra with positive and negative energies and, as argued in References \cite{JPA,Symm1}, this implies that one IR of the dS algebra describes a particle and its
antiparticle simultaneously. Since for the cosmological acceleration problem it is not necessary to consider 
antiparticles and spin effects, we give
only expressions for the action of the operators on the upper hyperboloid in the case of zero spin \cite{JPA,Symm1}:
\begin{eqnarray}
&&{\bf J}=l({\bf v}),\quad {\bf L}=-i v_0 \frac{\partial}{\partial {\bf v}},\quad {\bf B}=m_{dS} {\bf v}+i [\frac{\partial}{\partial {\bf v}}+
{\bf v}({\bf v}\frac{\partial}{\partial {\bf v}})+\frac{3}{2}{\bf v}]\nonumber\\
&& {\cal E}=m_{dS} v_0+i v_0({\bf v}
\frac{\partial}{\partial {\bf v}}+\frac{3}{2})
\label{IR1}
\end{eqnarray}
where ${\bf B}=\{M^{41},M^{42},M^{43}\}$, ${\bf l}({\bf v})=-i{\bf v}\times \partial/\partial {\bf
v}$, ${\cal E}=M^{40}$ and $m_{dS}$ is a positive quantity. 

This implementation of the IR is convenient for the transition to Poincare limit. Indeed, the operators
of the Lorenz algebra in Eq. (\ref{IR1}) are the same as in the IR of the Poincare algebra. Suppose
that the limit of $m_{dS}/R$ when $R\to\infty$ is finite and denote this limit as $m$. Then in the
limit $R\to\infty$ we get standard expressions for the operators of the IR of the Poincare algebra
where $m$ is standard mass, $E={\cal E}/R=mv_0$ and ${\bf P}={\bf B}/R=m{\bf v}$. For this
reason $m_{dS}$ has the meaning of the dS mass. 
Since Poincare symmetry is a special case of dS one, $m_{dS}$ is more fundamental than $m$.
Since Poincare symmetry works with a high accuracy, the value of $R$ is supposed to be very large (but, as noted above, it cannot be infinite).

Consider the non-relativistic approximation when $|{\bf v}|\ll
1$. If we wish to work with units where the dimension of
velocity is $meter/sec$, we should replace ${\bf v}$ by ${\bf
v}/c$. If ${\bf p}=m{\bf v}$ then it is clear from the
expression for ${\bf B}$ in Eq. (\ref{IR1}) that ${\bf p}$ becomes the real momentum ${\bf P}$
only in the limit $R\to\infty$. At this
stage we do not have any coordinate space yet. However, by analogy with
standard quantum mechanics, we can {\it define} the position
operator ${\bf r}$ as $i\partial/\partial {\bf p}$. 

In classical approximation we can treat
${\bf p}$ and ${\bf r}$ as usual vectors. Then as follows from Eq. (\ref{IR1})
\begin{equation}
{\bf P}= {\bf p}+mc{\bf r}/R, \quad H = {\bf p}^2/2m +c{\bf p}{\bf r}/R,\quad {\bf L}=-m{\bf r}
\label{PH}
\end{equation}
where $H=E-mc^2$ is the classical nonrelativistic Hamiltonian. As follows from these expressions, 
\begin{equation}
H({\bf P},{\bf r})=\frac{{\bf P}^2}{2m}-\frac{mc^2{\bf r}^2}{2R^2}
\label{HP}
\end{equation}

The last term in Eq. (\ref{HP}) is the dS correction to
the non-relativistic Hamiltonian. It is interesting to note
that the non-relativistic Hamiltonian depends on $c$ although
it is usually believed that $c$ can be present only in
relativistic theory. This illustrates the fact mentioned in Section \ref{fundamentaltheories}
that the transition to nonrelativistic theory
understood as $|{\bf v}|\ll 1$ is more physical than that
understood as $c\to\infty$. The presence of $c$ in Eq.
(\ref{HP}) is a consequence of the fact that this expression is
written in standard units. In nonrelativistic theory $c$ is
usually treated as a very large quantity. Nevertheless, the
last term in Eq. (\ref{HP}) is not large since we assume
that $R$ is very large.

As follows from Eq. (\ref{HP}) and the Hamilton equations, in dS theory a free particle
moves with the acceleration given by
\begin{equation}
{\bf a}={\bf r}c^2/R^2
\label{accel}
\end{equation}
where ${\bf a}$ and ${\bf r}$ are the acceleration and the radius vector of the particle, respectively.
Since $R$ is very large, the acceleration is not negligible only at cosmological distances
when $|{\bf r}|$ is of the order of $R$. 

Following our results in References \cite{JPA,Symm1}, we now consider whether the result (\ref{accel}) is compatible with GR. 
As noted in Subsec. \ref{CC}, the dS space is a four-dimensional manifold in the five-dimensional space defined by Eq. (\ref{dSspace}).
In the formal limit $R'\to\infty$ the action of the dS group in
a vicinity of the point $(0,0,0,0,x_4=R')$ becomes the action
of the Poincare group on Minkowski space. With this parameterization,
 the metric tensor on dS space is
\begin{equation}
g_{\mu\nu}=\eta_{\mu\nu}-x_{\mu}x_{\nu}/(R^{'2}+x_{\rho}x^{\rho})
\label{metric}
\end{equation}
where $\mu,\nu,\rho = 0,1,2,3$, $\eta_{\mu\nu}$ is the Minkowski metric tensor,  
and a summation
over repeated indices is assumed. It is easy to calculate the
Christoffel symbols in the approximation where all the
components of the vector $x$ are much less than $R'$:
$\Gamma_{\mu,\nu\rho}=-x_{\mu}\eta_{\nu\rho}/R^{'2}$. Then a
direct calculation shows that in the nonrelativistic
approximation the equation of motion for a single particle is
 the same as in Eq. (\ref{accel}) if $R'=R$.

Another way to show that Eq. (\ref{accel}) is compatible with GR follows. The known result of GR is that if the metric
is stationary and differs slightly from the Minkowskian one
then in the nonrelativistic approximation the curved space-time
can be effectively described by a gravitational potential
$\varphi({\bf r})=(g_{00}({\bf r})-1)/2c^2$. We now express
$x_0$ in Eq. (\ref{dSspace}) in terms of a new variable $t$ as
$x_0=t+t^3/6R^{'2}-t{\bf x}^2/2R^{'2}$. Then the expression for the
interval becomes
\begin{equation}
ds^2=dt^2(1-{\bf r}^2/R^{'2})-d{\bf r}^2-
({\bf r}d{\bf r}/R')^2
\label{II67}
\end{equation}
Therefore, the metric becomes stationary and $\varphi({\bf r})=-{\bf r}^2/2R^{'2}$ in agreement with Eq. (\ref{accel}) if $R'=R$. The fact that in classical approximation the parameter $R$ defining contraction from quantum dS
symmetry to quantum Poincare symmetry becomes equal the radius of dS space in GR does not mean
that $R$ can be always identified with this radius because on quantum level the notion of background space does not have a direct physical meaning.

Consider now a system of two free classical bodies in GR. Let $({\bf r}_i,{\bf a}_i)$
$(i=1,2)$ be their radius vectors and accelerations, respectively. Then Eq. (\ref{accel}) is
valid for each particle if $({\bf r},{\bf a})$ is replaced by $({\bf r}_i,{\bf a}_i)$, respectively.
Now if we define the relative radius vector ${\bf r}={\bf r}_1-{\bf r}_2$ and the 
relative acceleration ${\bf a}={\bf a}_1-{\bf a}_2$ then they will satisfy the same Eq. (\ref{accel})
which shows that the dS antigravity is repulsive. 

Let us now consider a system of two free bodies in the framework of the representation of the dS algebra. The particles are described by the
variables ${\bf P}_j$ and ${\bf r}_j$ ($j=1,2$). Define standard nonrelativistic variables
\begin{eqnarray}
&&{\bf P}_{12}={\bf P}_1+{\bf P}_2,
\quad {\bf q}=(m_2{\bf P}_1-m_1{\bf P}_2)/(m_1+m_2)\nonumber\\
&&{\bf R}_{12}=(m_1{\bf r}_1+m_2{\bf r}_2)/(m_1+m_2),\quad
{\bf r}={\bf r}_1-{\bf r}_2
\label{2body}
\end{eqnarray}
Then, as follows from Eq. (\ref{PH}), in the
nonrelativistic approximation the two-particle quantities ${\bf P}$, ${\bf E}$ and ${\bf L}$ are given by
\begin{equation}
{\bf P}= {\bf P}_{12},\quad E = M+\frac{{\bf P}_{12}^2}{2M} -\frac{Mc^2{\bf R}_{12}^2}{2R^2},
\quad {\bf L}=-M{\bf R}_{12}\label{2PE}
\end{equation}
where
\begin{equation}
M = M({\bf q},{\bf r})=
m_1+m_2 +H_{nr}({\bf r},{\bf q}),\quad 
H_{nr}({\bf r},{\bf q})=\frac{{\bf q}^2}{2m_{12}}-\frac{m_{12}c^2{\bf r}^2}{2R^2}
\label{2M}
\end{equation}
and $m_{12}$ is the reduced two-particle mass. Here the operator $M$ acts in the space of functions
$\chi({\bf q})$ such that $\int |\chi({\bf q})|^2d^3{\bf q}<\infty$ and ${\bf r}$ acts in this space as
${\bf r}=i\partial/\partial {\bf q}$. 

It now follows from Eq. (\ref{IR1}) that $M$ has the meaning of the two-body mass and
therefore $M({\bf q},{\bf r})$ is the internal two-body Hamiltonian. Then, by analogy with the derivation of Eq. (\ref{accel}),
it can be shown from the Hamilton equations that in semiclassical approximation the relative
acceleration is given by the same expression (\ref{accel}) but now
${\bf a}$ is the relative acceleration and ${\bf r}$ is the
relative radius vector.

This result has been obtained without using dS space, its
metric, connection etc.: it is simply a consequence of dS quantum mechanics of two free bodies
and {\it the calculation does not involve any geometry}. In our opinion this result is more important than the
result of GR because any classical result should be a consequence
of quantum theory in semiclassical approximation. Then, as follows from basic principles of quantum theory, correct description of nature in GR  
implies that $\Lambda$ {\it must} be nonzero, and the problem why $\Lambda$ is as is does not arise.
This has nothing to do with gravity, existence or nonexistence of dark energy and with the problem whether
or not  empty space-time should be necessarily flat. 

\section{What mathematics is more pertinent for describing nature?}
\label{whatmath}

In the preceding sections we discussed symmetries in standard quantum theory which is based on
classical mathematics. A belief of the overwhelming majority of scientists is that classical mathematics
(involving the notions of infinitely small/large and continuity) is the most general while finite mathematics 
is something inferior what is used only in special applications. This belief is based on the fact that
the history of mankind undoubtedly shows that classical mathematics has demonstrated its power in many areas of science. 

The notions of infinitely small, continuity etc. were proposed by Newton and Leibniz more than 300 years ago
and later were substantiated by Cauchy, Weierstrass, and Riemann. At that times people did not know about atoms and elementary particles. On the basis of everyday experience they believed that any macroscopic object can be divided into arbitrarily large number of arbitrarily small parts. However, from the point of view of the present knowledge those notions are problematic. For example, a glass of water contains approximately $10^{25}$ molecules.
We can divide this water by ten, million, etc. but when we reach the level of atoms and elementary particles
the division operation loses its usual meaning and we cannot obtain arbitrarily small parts. 

The discovery of atoms and elementary particles indicates that at the very basic level nature is discrete.
As a consequence, any description of macroscopic phenomena using continuity and differentiability can be only approximate.
For example, in macroscopic physics it is assumed that spatial coordinates and time are continuous measurable variables.
However, this is obviously an approximation because coordinates cannot be {\it directly} measured with the accuracy better than atomic
sizes and time cannot be measured with the accuracy better than $10^{-18}s$,
which is of the order of atomic size over $c$. 

As a consequence, distances less than atomic ones do not have a direct physical meaning and in real life there are no  
continuous lines and surfaces. As an example, water in the ocean can be described by differential equations of 
hydrodynamics but this is only an approximation since matter is discrete.  Another example is that if we draw a line on a sheet of paper 
and look at this line by a microscope then we will see that the line is strongly discontinuous because it consists of atoms.

Note that even the name "quantum theory" reflects a belief that nature is quantized, i.e.
discrete. Nevertheless, when quantum theory was created it was based on classical mathematics developed mainly in the 19th century. One of the greatest successes of the early quantum 
theory was the discovery that energy levels of the
hydrogen atom can be described in the framework of classical mathematics because the Schr\"{o}dinger 
differential operator has a discrete spectrum. This and many other successes of quantum theory were treated 
as indications that all   
problems of the theory can be solved by using classical mathematics. 

As a consequence, even after 90+ years of the existence of quantum theory it is still based on 
classical mathematics. Although the theory contains divergences and other inconsistencies, physicists persistently try to resolve them in the framework of classical mathematics. This situation is not natural but it is 
probably a consequence of historical reasons. The founders of quantum theory
were highly educated scientists but they used only classical mathematics, and even now discrete and finite mathematics is not a part of standard mathematical education at physics departments. Note that even regardless of
applications in physics, classical mathematics has its own foundational problems which cannot be resolved (as follows, in particular, from G\"{o}del's incompleteness theorems) and therefore the ultimate physical theory cannot be based on that
mathematics.  

In Section \ref{fundamentaltheories} we have formulated {\bf Definition} decribing when theory $A$ is more general
than theory $B$ and the latter is a special degenerated case of the former in the formal limit when a finite
parameter in the former goes to zero or infinity. In the subsequent sections we prove that the same {\bf Definition}
applies for the relation between finite quantum theory and finite mathematics on one hand, and standard
quantum theory and classical mathematics on the other. Namely, the former theories are based on a ring or field
with a finite characteristic $p$ and the latter theories are special degenerated cases of the former ones in the
formal limit $p\to\infty$.

In our publications (see e.g. References \cite{PRD,lev4,lev4B,lev4C,symm,lev4D}) we discussed an approach
called Finite Quantum Theory (FQT) where  quantum theory is based not on
classical but on finite mathematics. Physical states in FQT are elements of a linear space over a finite field or
ring, and operators of physical quantities are linear operators in this space. As noted in Section \ref{fundamentaltheories},
in standard quantum theory symmetry is defined by a Lie algebra of basic operators acting in the Hilbert space
of the system under consideration. Analogously, in FQT
symmetry is defined by a Lie algebra of basic operators acting in the space over a finite ring or field of
characteristic $p$. Following Reference \cite{monograph} we prove in Section \ref{FQT} that: a) FQT is more general than standard quantum theory and the latter is a special degenerated case of the former in the formal limit when the characteristic of the field or ring in FQT goes to infinity; b) finite mathematics itself  is more general than
classical mathematics and the latter is a special degenerated case of the former in the formal limit when the characteristic of the field or ring in finite mathematics goes to infinity.

\section{The problem of potential vs. actual infinity}
\label{infinity}

According to Wikipedia: "In the philosophy of mathematics, the abstraction of actual infinity involves the acceptance (if the axiom of infinity is included) of infinite entities, such as the set of all natural numbers or an infinite sequence of rational numbers, as given, actual, completed objects. This is contrasted with potential infinity, in which a non-terminating process (such as "add 1 to the previous number") produces a sequence with no last element, and each individual result is finite and is achieved in a finite number of steps.". 

The {\it technique} of classical mathematics involves only potential infinity, i.e. infinity is understood only as a limit
and, as a rule, legitimacy of every limit is thoroughly investigated.  However, {\it the basis} of classical mathematics does involve actual infinity: the infinite ring of integers $Z$ is the starting point for constructing infinite sets with different cardinalities, and, even in standard textbooks on classical mathematics, it is not even posed a problem whether $Z$ can be treated as a limit of finite sets. 

On the other hand, by definition, finite mathematics deals only with finite sets and finite numbers of elements (for example, in finitistic mathematics all natural numbers are considered but only finite sets are allowed). Known examples are theories of finite fields and finite rings described in a vast literature.
Finite mathematics starts from the ring $R_p=(0,1,...p-1)$ where all operations are modulo $p$. In the literature the notation $Z/p$ for $R_p$ is often used. We believe that this
notation is not quite consistent because it might give a wrong impression that finite mathematics starts from the
infinite set $Z$ and that $Z$ is more general than $R_p$. However, as proved in Section \ref{proof}, the situation is the opposite: although $R_p$ has less elements than $Z$, $R_p$ is more general than $Z$. 
Namely, in that section we prove  

{\bf Statement 1:} {\it The ring $Z$ is the limit of the ring $R_p$ when $p\to\infty$ since the result of any finite combination of additions, subtractions and multiplications in
$Z$ can be reproduced in $R_p$ if $p$ is chosen to be sufficiently large. On the contrary, when the limit is already taken then one cannot return back from $Z$ to
$R_p$, and in $Z$ it is not possible to 
reproduce all results in $R_p$ because in $Z$ there are no operations modulo a number.} 
Then, according to {\bf Definition} in Section \ref{fundamentaltheories}, {\it the ring $R_p$ is more general than $Z$, and $Z$ is a special degenerated case of $R_p$}.

In abstract mathematics there is no notion that one branch of mathematics
is more general than the other. For example,
classical and finite mathematics are treated as fully independent theories dealing with different problems. 
On the other hand, {\bf Definition} in Section \ref{fundamentaltheories}
describes conditions when one theory is more general than the other. 
A question arises whether {\bf Definition} can be used for proving that finite mathematics is more
general than classical one. As shown in Section \ref{FQT}, as a consequence of {\bf Statement 1},
quantum theory based on finite mathematics is more general than standard quantum theory based
on classical mathematics. Since quantum theory is the most general physical theory (all other
physical theories are special cases of quantum one), this implies that in applications finite mathematics is 
more pertinent  than classical one and that  

{\bf Main Statement: Even classical mathematics
itself is a special degenerated case of finite mathematics  in the formal limit when the characteristic of the field or ring in the latter goes to infinity}. 

As explained in Section \ref{FQT}, as a consequence,  theories with actual infinity can be only special degenerated cases of theories based on finite mathematics. We believe that to better understand the above problems it is important first to discuss in Section \ref{remarks} philosophical aspects of such a simple problem as operations with natural numbers.

\section{Remarks on arithmetic}
\label{remarks}

In the 20s of the 20th century the Viennese circle of philosophers
under the leadership of Schlick developed an approach called logical positivism which contains
verification principle:  {\it A proposition is only cognitively meaningful if it can be definitively and 
conclusively 
determined to be either true or false} (see e.g. References \cite{verificationism,verificationism1,verificationism2}). 
On the other hand, as noted by Grayling
\cite{Grayling}, {\it "The general laws of science are not, even in principle, verifiable, if verifying means 
furnishing conclusive proof of their truth. They can be strongly supported by repeated experiments and 
accumulated evidence but they cannot be verified completely"}. Popper proposed the 
concept of falsificationism \cite{Popper}: {\it If no cases where a claim is false can be found, then the hypothesis 
is accepted as provisionally true}. 

According to the principles of quantum theory, there should be no statements
accepted without proof and based on belief in their correctness (i.e. axioms).
The theory should contain only those statements that can be verified, at least in principle, 
where by "verified" physicists mean 
experiments involving only a finite number of steps. So the philosophy of quantum theory is similar to 
verificationism, not falsificationism. Note that Popper was a strong opponent of 
quantum theory and supported Einstein in his dispute with Bohr.

The verification principle does
not work in standard classical mathematics. For example, it cannot be determined whether the statement that 
$a+b=b+a$ for all natural numbers $a$ and $b$ is true or false. According
to falsificationism, this statement is provisionally true until one has found some numbers $a$ and $b$ for which $a+b\neq b+a$. There exist different theories of arithmetic
(e.g. finitistic arithmetic, Peano arithmetic or Robinson arithmetic) aiming to solve foundational problems of
standard arithmetic. However, those theories are incomplete and are not used in applications.

From the point of view of verificationism and principles of quantum theory, classical mathematics is not well defined
not only because it contains an infinite number of numbers. For example, let us pose a problem whether 
10+20 equals 30. Then one should describe an experiment
which gives the answer to this problem. Any computing device can operate only with a finite amount of resources and can perform
calculations only modulo some number $p$. Say $p=40$, then the experiment will confirm that
10+20=30 while if $p=25$ then one will get that 10+20=5. {\it So the statements that 10+20=30 and even that $2\cdot 2=4$
are ambiguous because they do not contain information on how they should be verified.} 
On the other hands, the statements
$$10+20=30\,(mod\, 40),\,\, 10+20=5\,(mod\, 25),$$
$$2\cdot 2=4\,(mod\, 5),\,\, 2\cdot 2=2\,(mod\, 2)$$
are well defined because they do contain such an information.
So, from the point of view of verificationism and principles of quantum theory, only operations modulo a number are well defined.

We believe the following 
observation is very important: although classical 
mathematics (including its constructive version) is a part of our everyday life, people typically do not realize that {\it classical mathematics is implicitly 
based on the assumption that one can have any desired amount of resources}. 
 In other words, standard operations with natural numbers are implicitly treated as limits of operations modulo $p$ 
when $p\to\infty$. Usually  in mathematics, legitimacy of every limit is thoroughly investigated, 
but in the simplest
case of standard operations with natural numbers it is not even mentioned that those
operations can be treated as limits of operations modulo $p$. In real life such limits even 
 might not exist if, for example, the Universe contains a finite number of elementary particles.

Classical mathematics proceeds from standard arithmetic which does not contain operations
modulo a number while finite mathematics necessarily involves such operations. In the
subsequent sections we explain that, regardless of philosophical preferences, finite mathematics
is more general than classical one. 

\section{Remarks on Statement 1}
\label{remarksonstatement1}

As noted above, {\bf Statement 1} is the first stage in proving that finite mathematics
is more general than classical one. Therefore this statement should not be based on
results of classical mathematics. In particular, it should not be based on properties
of the ring $Z$ derived in classical mathematics. The statement should be proved
by analogy with standard proof that a sequence of natural numbers
$(a_n)$ goes to infinity if $\forall M>0$ $\exists n_0$ such that $a_n\geq M\,\, \forall n\geq n_0$. In particular, the proof should involve only potential infinity but
not actual one.

The meaning of the statement is
that for any $p_0>0$ there exists a set $S$ and a natural number $n$ such that for any 
$m\leq n$ the result of any $m$ operations of multiplication, summation or subtraction of elements from $S$ is the same for any $p\geq p_0$ and that cardinality
of $S$ and the number $n$ formally go to infinity when $p_0\to\infty$. This means 
that for the set $S$ and number $n$ there is no
manifestation of operations modulo $p$, i.e. the results of any $m\leq n$ operations
of elements from $S$ are formally the same in $R_p$ and $Z$. 

In practice this means that if experiments involve only such sets $S$ and numbers $n$
then it is not possible to conclude whether the experiments are described by a
theory involving $R_p$ with a large $p$ or by a theory involving $Z$.

As noted above, classical mathematics starts from the ring $Z$, and, even in standard textbooks on classical mathematics, it is not even
posed a problem whether $Z$ can be treated as a limit of finite sets. We did not
succeed in finding {\it a direct proof} of {\bf Statement 1} in the literature.
However, the fact that $Z$ can be treated as a limit of $R_p$ when $p\to\infty$ follows from a sophisticated construction called ultraproducts. As shown e.g. in References \cite{ultraproducts,ultraproducts1},
infinite fields of zero characteristic (and Z) can be embedded in ultraproducts of finite fields. This fact can also be proved by using only rings (see e.g. Theorem 3.1 in
Reference \cite{Turner}). 
This is in the spirit of mentality of majority of mathematicians that sets with characteristic 0 are general, and for investigating those sets it is convenient to use properties of simpler sets of positive characteristics. 

The theory of ultraproducts (described in a wide literature --- see e.g. monographs
\cite{ultra,ultra1} and references therein) is essentially based on classical results on
infinite sets involving actual infinity. In particular, the theory is based on \L{}o\^s' theorem involving the
axiom of choice. Therefore theory of ultraproducts cannot be used in proving
that finite mathematics is more general than classical one.

Probably the fact that $Z$ can be treated as a limit of $R_p$ when $p\to\infty$, can also be proved in approaches not involving ultraproducts. For example, Theorem 1.1 in
Reference \cite{WooWoo} states:

{\it Let $S$ be a finite subset of a characteristic zero integral domain $D$, and let 
$L$ be a finite set of non-zero elements in the subring $Z[S]$ of $D$. There exists an infinite sequence of primes with positive relative density such that for each prime $p$ in the sequence, there is a ring
homomorphism $\varphi_p$ : $Z[S] \to Z/pZ$ such that 0 is not in $\varphi_p(L)$.}

The theorem involves only primes, and the existence of homomorphism does not
guarantee that operations modulo $p$ are not manifested for a sufficient number
of operations. However, even if those problems can be resolved, the proof of
the theorem is based on the results of classical mathematics for characteristic zero
integral domains, and the proof involves real and complex numbers, i.e. the results involve
actual infinity. 

We conclude that the existing proofs that $Z$ can be treated as a limit of $R_p$ when 
$p\to\infty$ cannot be used in the proof that finite mathematics is more general than classical one. 

\section{Proof of Statement 1}
\label{proof}

Since operations in $R_p$ are modulo $p$, one can represent 
$R_p$ as a set $\{0,\pm 1,\pm 2,...,\pm(p-1)/2)\}$ if $p$ is odd and as a set
$\{0,\pm 1,\pm 2,...,\pm (p/2-1),p/2\}$ if $p$ is even. Let $f$ be a function 
from $R_p$ to $Z$ such that
$f(a)$ has the same notation in $Z$ as $a$ in $R_p$. 
If elements of $Z$ are depicted as integer points on the $x$ axis of the $xy$ plane then, if $p$
is odd, the elements of $R_p$
can be depicted  as points of the circumference in Figure 1
\begin{figure}[!ht]
\centerline{\scalebox{0.3}{\includegraphics{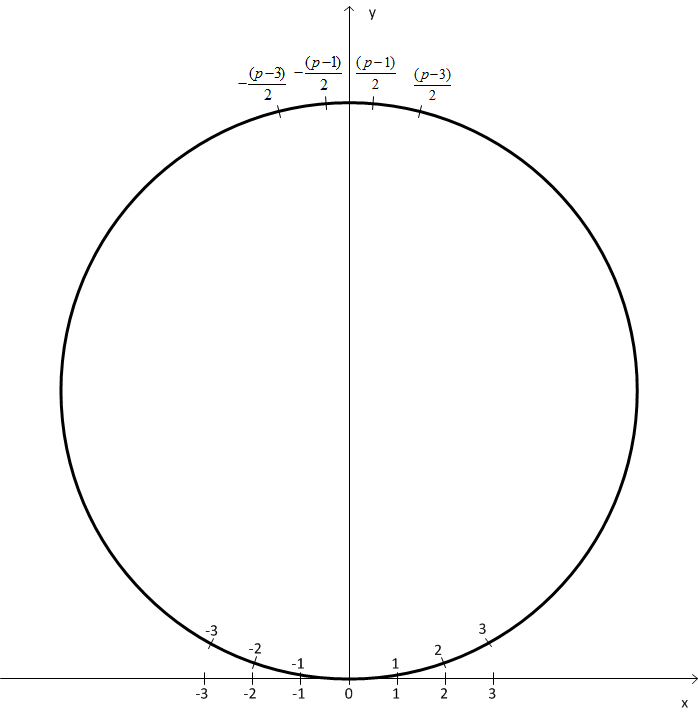}}}
\caption{
  Relation between $R_p$ and $Z$
}
\label{Fig.1}
\end{figure}
and analogously if $p$ is even. 
This picture is natural  since $R_p$ has a property that if we take any element $a\in R_p$ and sequentially add 1 then after $p$ steps we will exhaust
the whole set $R_p$ by analogy with the property that if we move along a circumference in the same direction
then sooner or later we will arrive at the initial point. 

We define the function $h(p)$ such that $h(p)=(p-1)/2$ if $p$ is odd and $h(p)=p/2-1$ if $p$ is even.
Let $n$ be a natural number and $U(n)$ be a set of elements $a\in R_p$ such that 
$|f(a)|^n \leq h(p)$. Then $\forall m \leq n$ the result of any $m$ operations of addition,
subtraction or multiplication of elements $a\in U(n)$ is the same as for the corresponding elements $f(a)$ in $Z$, i.e. in this
case operations modulo $p$ are not explicitly manifested.

Let $n=g(p)$ be a function of $p$ and $G(p)$ be a function such that the set $U(g(p))$ contains
the elements $\{0,\pm 1,\pm 2,..., \pm G(p)\}$. In what follows $M>0$ and $n_0>0$ are natural numbers.
If there is a sequence of natural numbers
$(a_n)$ then standard definition that $(a_n)\to \infty$ is that $\forall M$ $\exists n_0$ 
such that $a_n\geq M\,\, \forall n\geq n_0$. By analogy with this
definition we will now prove 

{\it Proposition: There exist functions $g(p)$ and $G(p)$ such that $\forall M$ $\exists p_0>0$ such that $g(p)\geq M$ and $G(p)\geq 2^M\,\, \forall p\geq p_0$.}
 
\begin{proof}
$\forall p>0$ there exists a unique natural $n$ such that $2^{n^2}\leq h(p)<2^{(n+1)^2}$. Define
$g(p)=n$ and $G(p)=2^n$. Then $\forall M\,\, \exists p_0$ such that $h(p_0)\geq 2^{M^2}$. Then
$\forall p\geq p_0$ the conditions of {\bf Statement 1} are satisfied.
\end{proof}

As a consequence of {\it Proposition} and {\bf Definition}, {\bf Statement 1}  is valid, i.e. 
the ring $Z$ is the limit of the ring $R_p$ when $p\to\infty$ and $Z$ is a special degenerated case of $R_p$.

When $p$ is very large then $U(g(p))$ is a relatively small part of $R_p$, and in general the results
in $Z$ and $R_p$ are the same only in $U(g(p))$. This is analogous to the fact mentioned in
Section \ref{fundamentaltheories} that the results of NT and RT are the same only in relatively small cases
when velocities are much less than $c$. However, 
 when the radius of the circumference in Figure 1 becomes infinitely large then
a relatively small vicinity of zero in $R_p$ becomes the infinite set $Z$ when $p\to\infty$. 
{\it This example demonstrates that once we involve actual infinity and replace $R_p$ by $Z$ then we automatically 
obtain a degenerated theory because in $Z$ there are no operations modulo a number}.

\section{Why finite mathematics is more general than classical one}
\label{FQT}

As noted in Section \ref{infinity}, finite mathematics is more general than classical one if 
finite mathematics is more pertinent in applications than classical one. Since quantum theory is the most
general physical theory (i.e. all other physical theories are special cases of quantum one), 
the answer to this question depends on whether standard quantum theory
based on classical mathematics is the most general or is a special degenerated case of a more general 
quantum theory.

In classical mathematics, the ring $Z$ is the starting point for introducing the notions of rational, real, complex 
 numbers etc. Therefore those notions arise from a degenerated set. Then a question arises whether the fact that $R_p$ is more general than Z implies
that finite mathematics is more general than classical one, i.e. whether finite 
mathematics can reproduce all results obtained by applications of classical mathematics.
For example, if $p$ is prime then $R_p$ becomes the Galois field $F_p$, and the results
in $F_p$ considerably differ from those in the set $Q$ of rational numbers even when $p$ is
very large. In particular, 1/2 in 
$F_p$ is a very large number $(p + 1)/2$.  

As proved in Section \ref{fundamentaltheories}, standard dS/AdS quantum theories are more general than standard
Poincare quantum theory. In the former, quantum states are described by representations of the
dS/AdS algebras. To make relations between standard quantum
theory and FQT more straightforward, we will modify the commutation relations (\ref{CR}) 
by writing them in the form 
\begin{equation}
[M^{ab},M^{cd}]=-2i (\eta^{ac}M^{bd}+\eta^{bd}M^{ac}-
\eta^{ad}M^{bc}-\eta^{bc}M^{ad})
\label{newCR}
\end{equation}
One might say that these relations are written in units $\hbar/2=c=1$. However, as noted in Section
\ref{fundamentaltheories}, quantum theory itself does not involve quantities $\hbar$ and $c$ at all, and Eq. (\ref{newCR})
indeed does not contain those quantities. The reason for writing the commutation relations in the form (\ref{newCR})
rather than (\ref{CR}) is that in this case the minimum nonzero value of the angular momentum is 1 instead
of 1/2. Therefore the spin of fermions is odd and the spin of bosons is even. This is convenient in FQT where, as noted above,  1/2 is a very large number. 

According to principles of quantum theory, from these ten operators one should construct a maximal set $S$
of mutually commuting operators defining independent physical quantities and construct a basis in the
representation space such that the basis elements are eigenvectors of the operators from $S$.

The rotation subalgebra of algebra (\ref{newCR}) is described in every textbook on quantum mechanics. 
The basis of the subalgebra is $(J_x,J_y,J_z)=(M^{23},M^{31},M^{12})$, and with the choice of the commutation relations in the form of Eqs. (\ref{newCR})  instead of Eq. (\ref{J}), the commutation relations between those  
operators operators are 
\begin{equation}
[J_x,J_y]=2iJ_z,\quad [J_z,J_x]=2iM_y,\quad [J_y,J_z]=2iJ_x
\label{MxMy}
\end{equation}
A possible choice of $S$ is $S=(J_z,K)$ where $K=J_x^2+J_y^2+J_z^2$ is the Casimir
operator of the subalgebra, i.e. it commutes with all the operators of the subalgebra. Then any irreducible
representation of the subalgebra is described by an integer $k\geq 0$. The basis elements $e(\mu , k)$ of the representation space are eigenvectors of the operator $K$ with the eigenvalue $k(k+2)$ and
the eigenvectors of the operator $J_z$ with the eigenvalues $\mu$ such that, for a given $k$, $\mu$ can
take $k+1$ values $\mu=-k,-k+2,...,k-2,k$. Therefore all the basis elements are eigenvectors of the operators
from $S$ with the eigenvalues belonging to $Z$.

In Sections. 4.1 and 8.2 of Reference \cite{monograph} we discussed the dS and AdS cases, 
respectively and have shown that

{\bf Statement 2:} {\it For algebra (\ref{newCR}) there exist sets $S$ and representations such that basis vectors
in the representation spaces are eigenvectors of the
operators from $S$ with eigenvalues belonging to $Z$. Such representations reproduce
standard representations of the Poincare algebra in the formal limit $R\to\infty$}.
Therefore the remaining problem is whether or not FQT  can be a generalization of standard quantum theory where
states are described by elements of a separable complex Hilbert spaces $H$. 

FQT can be defined as a quantum theory where states are elements
of a space over a finite ring or field with characteristic $p$ and operators of physical quantities act in this space. 
In FQT symmetry can be defined in full analogy with standard quantum theory. In particular, dS and AdS
symmetries can be defined such that the commutation relations between basic operators $M^{ab}$ are
the same as in Eq. (\ref{newCR}). 

Let $x$ be an element of $H$ and $(e_1,e_2,...)$ be
a basis of $H$ normalized such that the norm of each $e_j$ is an integer. Then {\it with any desired 
accuracy each element of $H$ can be approximated by a finite linear combination
\begin{equation}
x=\sum_{j=1}^n c_je_j 
\label{sum}
\end{equation}
where $c_j=a_j+ib_j$ and all the numbers $a_j$ and $b_j$ ($j=1,2,....n$) are
rational}. 
This follows from the known fact that the set of such sums is dense in $H$.

The next observation is that spaces in standard quantum theory are projective, i.e. for any complex
number $c\neq 0$ the elements $x$ and $cx$ describe the same state. This follows from the
physical fact that not the probability itself but only ratios of probabilities have a physical
meaning. In view of this property, both parts of Eq. (\ref{sum}) can be multiplied by a 
common denominator of all the numbers $a_j$ and $b_j$. As a result, we have

{\bf Statement 3:} {\it Each element of $H$ can be approximated by a finite linear combination (\ref{sum})
where all the numbers $a_j$ and $b_j$ belong to $Z$}.

In the literature it is also considered a version of quantum theory based not on real but on $p$-adic numbers (see e.g. the review paper \cite{padic} and references therein). 
Both, the sets of real and $p$-adic numbers
are the completions of the set of rational numbers but with respect to different metrics. Therefore the set of rational numbers is dense in both, in the set of real numbers and in
the set of $p$-adic numbers ${\cal Q}_p$. In the $p$-adic case, the Hilbert space analog of $H$ is the space of complex-valued functions $L^2({\cal Q}_p)$ and therefore there is an analog of {\bf Statement 3}.

We conclude that Hilbert spaces in standard quantum theory contain a big redundancy of
elements. Indeed, although formally the description of states in standard quantum theory
involves rational and real numbers, such numbers play only an auxiliary role because 
with any desired accuracy each state can be described by using only integers. 
Therefore, as follows from {\bf Definition} in Section. \ref{fundamentaltheories} and {\bf Statements 1-3},
\begin{itemize}
\item Standard quantum theory based on classical mathematics is a special degenerated case
of FQT in the formal limit $p\to\infty$.
\item {\bf Main Statement} formulated in Section \ref{infinity} is valid.
\end{itemize}

\section{Discussion}
\label{conclusion}

In Section \ref{fundamentaltheories} we argue that in quantum theory symmetry is defined by a Lie algebra
of basic operators. In the theory of Lie algebras there exist a clear criterion when the Lie algebra $A_1$
is more general than the Lie algebra $A_2$: when $A_2$ can be obtained from $A_1$ by contraction.
As a consequence, dS and AdS quantum theories are more general
than Poincare quantum theory. We note that the same conclusion has been given in the famous
Dyson's paper \cite{Dyson} where symmetries were treated in terms of Lie groups rather than Lie algebras.

The paper \cite{Dyson} appeared in 1972 and, in view of Dyson's results, a question arises why general theories
of elementary particles (QED, electroweak theory and QCD) are still based on Poincare symmetry and not dS or AdS
symmetries. Probably a justification is that since the parameter of contraction $R$ from dS or AdS theories to Poincare one is much greater that sizes of elementary particles, there is
no need to use the former symmetries for description of elementary particles.

We believe that this argument is not consistent because usually more general theories shed a new
light on standard concepts. For example, as shown in References \cite{JPA,Symm1}, in contrast to the
situation in Poincare invariant theories, where a particle and its antiparticle are described by different
irreducible representations (IRs) of the Poincare algebra (or group), in dS theory a particle and its antiparticle
belong to the same IR of the dS algebra. In the formal limit $R\to\infty$ one IR of the dS algebra splits into two
{\it different} IRs of the Poincare algebra for a particle and its antiparticle. Strictly speaking, this implies that in dS
theory the very notion of a particle and its antiparticle is only approximate since transitions 
particle$\leftrightarrow$antiparticle are not strictly prohibited. As a consequence, in dS theory 
the electric charge and the baryon and lepton quantum numbers are only approximately conserved. 
At present they are conserved with a high
accuracy. However, one might think that at early stages of the Universe the quantity $R$ was much less
than now and the nonconservation of those quantum numbers was much stronger. This might be a
reason of the known phenomenon of baryon asymmetry of the Universe.

Physicists usually understand that physics cannot (and should not) derive that $c\approx 3\cdot 10^8m/s$ 
and $\hbar \approx 1.054\cdot 10^{-34}kg\cdot m^2/s$. At the same time they usually believe that physics should derive the value of the cosmological constant $\Lambda$ and that the solution of the dark energy problem depends on this value. As noted in Section \ref{fundamentaltheories}, the question why the quantities $(c,\hbar)$ are as are does not arise since the
answer is: because people's choice is to measure velocities in $m/s$ and angular momenta in $kg\cdot m^2/s$.
In the modern system of units it is postulated that the quantities $(c,\hbar)$ are the same at all times.
At the same time, as noted in Sections. \ref{intro} and \ref{fundamentaltheories}, background space has 
a clear physical meaning only on classical level while on quantum one transition from dS/AdS symmetries to Poincare symmetry is
characterized by the parameter $R$ which is the coefficient of proportionality between the quantities $M^{4\mu}$ and $P^{\mu}$ and has nothing to do with the radius of background space. This parameter {\it must} be finite and, as noted in Section. \ref{fundamentaltheories}, 
 the question why $R$ is as is does not arise because people's choice
is to measure distances in meters. 
Since the modern system of units is based on Poincare invariance, it says nothing on whether $R$ can change with time. This is the problem of metrology and cosmology but not fundamental physics.

Therefore the quantity $\Lambda$ is meaningful only on classical level. As shown in Section \ref{acceleration}, as a consequence of dS symmetry on quantum level, in semiclassical approximation two free bodies have a relative acceleration given by the same expression as in GR if the the radius of dS space equals $R$ and $\Lambda=3/R^2$. 
We believe that our result is more important than the result of GR for the following reasons.

In GR there is no restriction on the radius of background space and, for example, the possibility $\Lambda=0$
is not excluded. As noted in Subsec. \ref{CC}, this possibility is now adopted by majority of physicists and,
as a consequence, the results on cosmological acceleration are interpreted as a consequence of the existence
of dark energy. However, from the point of view of symmetry on quantum level discussed in the present
paper, dS symmetry is more general than Poincare one and there is no freedom in choosing the value of $R$.
Our result for the cosmological acceleration has been obtained without using dS space, its
metric, connection etc.: it is simply a consequence of dS quantum mechanics of two free bodies
and {\it the calculation does not involve any geometry}. The result is more important than the
result of GR because any classical result should be a consequence
of quantum theory in semiclassical approximation. The result has nothing to do with gravity, existence or nonexistence of dark energy and with the problem whether or not  empty space-time should be necessarily flat. 
Therefore, as noted in Section \ref{acceleration}, the cosmological constant problem and dark energy problem do not arise.

As noted in Section \ref{acceleration}, our conclusion is based on the consideration of the pure mathematical
problem: we consider quantum theory where dS symmetry is understood as explained in Section
\ref{fundamentaltheories} and then we consider semiclassical approximation in this theory. 
On the other hand, as noted in Section \ref{acceleration}, there exists a wide literature where standard 
GR is modified and,
as a consequence, the existing data can be explained by using alternatives to standard understanding
of dark energy. Therefore
the explanation of the of the data is a difficult unsolved problem, and we believe that for solving this problem
different approaches should be taken into account.

As argued in Section \ref{whatmath}, at the very general level,  quantum theory should not be based on
classical mathematics involving the notions of infinitely small/large, continuity, differentiability etc. In Section
\ref{infinity} we give a new look at the problem of potential vs. actual infinity. In classical mathematics the
infinite ring of integers $Z$ is the starting point for introducing the notions of rational, real, complex
number and other infinite sets with different cardinalities. However, in Section \ref{proof} we give 
a {\it direct} proof of {\bf Statement 1} that $Z$ is a
special degenerated case of the ring $R_p=(0,1,...p-1)$ in the formal limit $p\to\infty$, and the proof
does not involve actual infinity. 
We did not succeed in finding such a proof in the literature and, as noted above, even in standard
textbooks on classical mathematics, it is not even posed a problem whether $Z$ can be treated as a limit of finite sets. 
As noted in Section \ref{remarksonstatement1}, the fact that $Z$ can be treated as a limit of 
$R_p$ in the formal limit $p\to\infty$ can be proved proceeding from ultraproducts and other
sophisticated approaches. However, those approaches involve actual infinity and therefore they cannot be
used in the proof that finite mathematics is more general than classical one.

Note also, that the phrase that $Z$ is the ring of characteristic 0 reflects the usual spirit that $Z$ is more general than $R_p$. In our opinion it is natural to say that $Z$ is the ring of characteristic $\infty$ because it is a limit of rings of characteristics $p$ when
$p\to\infty$. The characteristic of the ring $p$ is understood such that all operations in the ring are modulo $p$ but operations modulo 0 are meaningless. Usually the characteristic of the ring is defined as the smallest positive number $n$ such that the sum of $n$ units $1+1+1...$ in the ring equals zero if such a number $n$ exists and 0 otherwise. However,  this sum can be written as $1\cdot n$ and the equality $1\cdot 0=0$ takes place in any ring.

Legitimacy of the limit of $R_p$ when $p\to\infty$ is problematic because when $R_p$ is replaced by $Z$
which is used as the starting point for constructing classical mathematics, we get classical
mathematics which has foundational problems. For example, G\"{o}del's incompleteness theorems state that no system of axioms can ensure that all facts about natural numbers can be proven and the system of axioms in classical mathematics cannot demonstrate its own consistency. The foundational problems of classical mathematics
arise as a consequence of the fact that the number of natural numbers is infinite. On the other
hand, since finite mathematics deals only with a finite number of elements, it does not have foundational
problems because here every statement can be directly verified, at least in principle. 

The efforts of many great mathematicians to resolve foundational problems of classical mathematics
have not been successful yet. The philosophy of Cantor, Fraenkel, G\"{o}del, Hilbert, Kronecker, Russell, Zermelo and other great mathematicians was based on macroscopic experience in which the 
notions of infinitely small, infinitely large, continuity and standard division are natural. 
However, as noted above, those notions contradict the existence of elementary particles and are not natural
in quantum theory. The illusion of continuity arises when one neglects discrete structure of matter.

The above construction has a known historical analogy. For many years people believed
that the Earth was flat and infinite, and only after a long period of time they realized that
it was finite and curved. It is difficult to notice the curvature dealing only with
distances much less than the radius of the curvature. Analogously one might think that
the set of numbers describing nature in our Universe has a "curvature" defined by a very
 large number $p$ but we do not notice it dealing only with numbers much less than $p$. 

As noted in Section \ref{FQT}, by analogy with standard quantum theory, one can construct a finite quantum theory (FQT) where the space of possible states of the system under consideration is a linear space over a finite ring or field
of characteristic $p$, and the operators of physical quantities are linear operators in that space.
By analogy with standard quantum theory, symmetry in FQT is defined by a Lie algebra of basic
operators. In particular, dS or AdS symmetry in FQT means that the structure constants of this Lie
algebra is the same as in standard case. Note that in FQT there can be no system of units, no infinitely small/large quantities and no continuity because all physical quantities are elements of a finite ring or field. 

In particular, FQT does not contain infinities at all and all operators are
automatically well defined. In my discussions with physicists,
some of them commented this fact as follows. This is an
approach where a cutoff (the characteristic $p$ of the finite ring or 
field) is introduced from the beginning and for this reason
there is nothing strange in the fact that the theory does not
have infinities. It has a large number $p$ instead and this
number can be practically treated as infinite. 

The inconsistency of this argument is clear from the following analogy. 
It is not correct to say that relativistic theory is simply nonrelativistic one with the cutoff $c$ for
velocities. As a consequence of the fact that $c$ is finite, relativistic theory considerably differs from nonrelativistic one in several aspects.
The difference between finite rings or fields on one hand 
and usual complex numbers on the other
is not only that the former are finite and the latter are
infinite. If the set of usual numbers is visualized as a
straight line from $-\infty$ to $+\infty$ then the simplest
finite ring can be visualized not as a segment of this line
but as a circumference (see Figure \ref{Fig.1} in Section \ref{proof}). This
reflects the fact that in finite mathematics the rules of arithmetic
are different and, as a result, FQT has many unusual features
which have no analogs in standard theory.

As noted in Section \ref{infinity}, in abstract mathematics there is no notion that one branch of mathematics
is more general than the other. However, as proved in Section \ref{FQT}, in applications finite mathematics is 
more pertinent  than classical one because FQT is more general than standard quantum theory:
the latter is a special degenerated case of the former in the formal limit when the characteristic of the ring
or field in the former goes to infinity. Therefore, the problem what branch of mathematics is more
general is the problem of physics, not mathematics and, as formulated in {\bf Main Statement}, finite mathematics
is more general than classical one. 

The fact that at the present stage of the Universe $p$ is a huge number explains why in many cases classical mathematics describes natural phenomena with a very high accuracy. 
At the same time, as shown in References \cite{PRD,symm,monograph}, the explanation of several phenomena can be given only in the theory where $p$ is finite.

One of the examples is that in our approach gravity is a manifestation of the fact that $p$ is finite.
In Reference \cite{monograph} we have derived the approximate expression for the gravitational constant which depends
on $p$ as $1/lnp$. By comparing this expression with the experimental value we get that
$lnp$ is of the order of $10^{80}$ or more, i.e. $p$ is a huge number of the order of $exp(10^{80})$ or more. However, since $lnp$ is "only" of the order of
$10^{80}$ or more, the existence of $p$ is observable while
in the formal limit $p\to\infty$ gravity disappears. 

{\bf Funding:} This research received no external funding.

 {\it Acknowledgements:} I am grateful to Bernard Bakker, Jos{\'e} Manuel Rodriguez Caballero, Vladimir Karmanov, Dmitry Logachev, Harald Niederreiter, Metod Saniga  and Teodor Shtilkind for numerous important discussions and
to Efim Zelmanov for telling me about ultraproducts and References \cite{ultraproducts,ultraproducts1}.

{\bf Conflicts of Interest:} The authors declare no conflict of interest.

\end{document}